\documentclass[9pt,final,journal]{IEEEtran}
\usepackage{amsmath,amssymb}
\usepackage{enumerate}
\usepackage{xcolor}
\newtheorem{teo}{Theorem}[section]
\newtheorem{defin}[teo]{Definition}
\newtheorem{lema}[teo]{Lemma}
\newtheorem{rem}[teo]{Remark}
\newtheorem{cor}[teo]{Corollary}

\newtheorem{example}{Example}

\def\K{\mathcal{K}}
\def\Ki{\K_{\infty}}
\def\KL{\mathcal{KL}}
\def\N{{\mathbb N}}
\def\R{{\mathbb R}}
\def\U{\mathcal{U}}

\def\mer{$\hfill\circ$}

\def\I{\mathcal{I}}

\title{(Integral-)ISS of switched and time-varying impulsive systems\\ based on global state weak linearization}

\author{J.L. Mancilla-Aguilar and H. Haimovich
\thanks{J.L. Mancilla-Aguilar is with Departamento de Matem\'atica, Instituto Tecnol\'ogico de Buenos Aires, Av. E. Madero 399, Buenos Aires, Argentina. \texttt{jmancill@itba.edu.ar}}
 \thanks{H. Haimovich is with CIFASIS, CONICET-UNR, Ocampo y Esmeralda, 2000 Rosario, Argentina. {\texttt{haimovich@cifasis-conicet.gov.ar}}}%
\thanks{Work partially supported by FONCyT grant PICT 2018-01385, Argentina.}%
}

\begin{document}
\maketitle


\begin{abstract}
 It is shown that impulsive systems of nonlinear, time-varying and/or switched form that allow a stable global state weak linearization are jointly input-to-state stable (ISS) under small inputs and integral ISS (iISS). The system is said to allow a global state weak linearization if its flow and jump equations can be written as a (time-varying, switched) linear part plus a (nonlinear) pertubation satisfying a bound of affine form on the state. This bound reduces to a linear form under zero input but does not force the system to be linear under zero input. The given results generalize and extend previously existing ones in many directions: (a) no (dwell-time or other) constraints are placed on the impulse-time sequence, (b) the system need not be linear under zero input, (c) existence of a (common) Lyapunov function is not required, (d) the perturbation bound need not be linear on the input.
\end{abstract}



\section{Introduction}
\label{sec:introduction}
Systems with time-driven impulse effects, also called impulsive systems or systems with state resets, have interesting stability properties caused by the interplay between the continuous (between impulses) and the discontinuous (at the impulse time instant) evolutions \cite{lakbai_book89,samper_book95}. The continuous evolution is defined by an ordinary differential equation ---the flow equation--- and the discontinuity occurring at an impulse instant by an algebraic equation ---the jump equation. External inputs may thus affect both the flow and the jump equations, and input-to-state stability (ISS) and integral ISS (iISS)-related properties \cite{sontag_tac89,sontag_scl98} must take this fact into account \cite{heslib_auto08}.

ISS and iISS results for impulsive systems based on Lyapunov-type functions abound. These exist for impulsive systems involving
switching \cite{liuliu_scl12,gaowan_nahs16}, interconnections \cite{daskos_nahs12,dasfek_nahs17}, sampling \cite{zhatan_auto19}, and also for specific types of infinite-dimensional systems \cite{dasmir_siamjco13,mirpri_siamrev20}. {\color{blue} The results for hybrid systems in \cite{caitee_scl09,norkha_mcss17} can also be helpful for establishing the ISS/iISS of impulsive systems.}

Lyapunov functions are a very useful tool for stability and robustness analysis. 
However, when some stability property is required to hold uniformly for every system in a given family of systems (e.g. switched systems under constrained switching) a single Lyapunov function common to all the systems in the family need not exist or may be very difficult to find.
Fortunately, some characterizations of the iISS property can be established without the aid of Lyapunov functions \cite{haiman_tac18}, and hence some well-known properties and relationships between ISS and iISS remain valid for families of time-varying and switched systems \cite{haiman_auto18}, even in the impulsive case \cite{haiman_aadeca18}.

Most existing asymptotic stability-related results for impulsive systems, including those already mentioned, ensure some type of convergence as continuous time advances without explicitly accounting for the number of impulses occurring within the elapsed time. This notion of asymptotic stability is weak in the sense that it allows for no meaningful robustness unless constraints are placed on the impulse-time instants \cite{haiman_auto20b}. Requiring convergence not only as continuous time advances but also as the number of occurring impulses increases {\color{blue}(as is standard in the context of hybrid systems \cite{goesan_book12})} leads to a stronger notion of stability that allows to recover, for impulsive systems, many of the robustness properties exhibited by nonimpulsive systems \cite{haiman_auto20,manhai_lcss20}. Moreover, this stronger stability notion becomes equivalent to the weak and most usual one whenever the number of occurring impulses can be bounded in correspondence with the length of the time interval over which they occur \cite[Proposition~2.3]{manhai_tac20}; this happens, e.g., when the continuous evolution (i.e. between impulses) has a minimum, average or fixed dwell time. 

Stability results for impulsive systems without inputs based on a first-order approximation have been known for some time \cite[Theorem~37]{samper_book95}, \cite{fenpin_na99}. In the presence of inputs, however, very few works provide this type of result.
In \cite[Theorem~3.2]{liudou_scl14}, it is shown that a linear time-varying impulsive system admits linear perturbations without losing its exponential stability, nor its exponential ISS with respect to additive inputs, provided a minimum dwell-time constraint is imposed between the occurrence of impulses. 
%
%
Bilinear systems \cite{mohler_tssc70}, where the dynamic equations may contain cross products between inputs and states, have been shown to have some interesting properties in the time-invariant and nonimpulsive case. Sontag \cite[Theorem~5]{sontag_scl98} established that asymptotic stability under zero input is actually equivalent to iISS for time-invariant and nonimpulsive systems. Moreover, \cite[Section~IV-C]{chaang_tac14} strengthened this result by showing that not just iISS but also ISS under small inputs follows from asymptotic stability under zero input. The fact that asymptotic stability under zero input implies iISS is extended to specific types of bilinear infinite-dimensional systems in \cite{mirito_mcrf16}.

The main contribution of the current paper is to provide ISS and iISS results based on linearization-type arguments that generalize existing results in several directions. Specifically, families of nonlinear impulsive systems with time-varying flow and jump maps are considered. It is shown that if the maps of every system in a given family can be written in a linear-plus-perturbation form, where the linear part is stable and the perturbation part admits a bound of specific form, then the family of systems is uniformly iISS, ISS or ISS with respect to small inputs, depending on the properties of the bound. Systems that admit this decomposition are said to admit a global state weak linearization. Such systems need not be linear under zero input, nor linear in the input; this certainly covers bilinear systems but also allows far greater generality.

\section{Preliminaries}

\subsection{Single impulsive system}
Consider an impulsive system $\Sigma$ of the form
\begin{subequations}
  \label{eq:nlsys}
  \begin{align}
    \label{eq:flow}
    \dot{x}(t) &= f(t,x(t),u(t)), & t &\notin\sigma,\\
    \label{eq:jump}
    x(t) &= 
    g(t,x(t^-),u(t)), & t &\in \sigma,
  \end{align}
\end{subequations}
{\color{blue}
where for $t\ge 0$, $x(t)\in \R^n$, $u(t)\in \R^m$ and $\sigma\in \Gamma$, where $\Gamma$ is the set of all sequences $\sigma=\{\tau_k\}_{k=1}^N$, with $N\in \N$ or $N=\infty$, satisfying $0<\tau_1<\tau_2<\cdots$ and $\lim_{k\to \infty}\tau_k=\infty$ when $N=\infty$. Depending on the context, $\sigma$ may also be regarded as the set which contains only the elements of the sequence.}

A locally bounded and Lebesgue measurable function $u:\R_{\ge 0}\to \R^m$ is called an input and $\U$ denotes the set of all inputs. A solution to (\ref{eq:nlsys}) is a right-continuous function $x : [t_0,T_x) \to \R^n$ that is locally absolutely continuous on each interval $J=[a,b)$ such that $(a,b)\cap \sigma=\emptyset$, its discontinuities are of first kind and satisfies~(\ref{eq:flow}) for almost all $t\in [t_0,T_x) \setminus \sigma$ and (\ref{eq:jump}) for all $t\in (t_0,T_x) \cap \sigma$ (see, e.g.~\cite{manhai_lcss20} for the precise definition of solution). Note that a solution always begins by flowing and not by jumping, even if $t_0\in\sigma$.

\subsection{Families of systems and basic assumptions}

The bounds on the trajectory ensured by stability or other properties are many times required to hold not just for a single system but uniformly for every system in a given family. The most usual case when this happens is in the study of switched systems, where each possible trajectory corresponds to a time-varying system but the interest is placed on all the possible trajectories that may occur for every admissible switching evolution. In order to cover these and other cases, we consider a parametrized family $\{\Sigma_\nu : \nu \in \Lambda\}$ of impulsive systems of the form (\ref{eq:nlsys}),
\begin{subequations}
  \label{eq:nlsys-f}
  \begin{align}
    \label{eq:flow-f}
    \dot{x}(t) &= f_\nu(t,x(t),u(t)), & t &\notin\sigma_\nu,\\
    \label{eq:jump-f}
    x(t) &= 
    g_\nu(t,x(t^-),u(t)), & t &\in \sigma_\nu,
  \end{align}
\end{subequations}
where $\sigma_\nu = \{\tau_k^\nu\}_{k=1}^{N} \in \Gamma$, $\nu$ is a parameter that identifies a single system within the considered family and $\Lambda$ is an arbitrary nonempty set, usually having an infinite and uncountable number of elements {\color{blue} (see Section \ref{sec:STVIS}, and ~\cite[Section~2.3]{haiman_auto19} and \cite[Section~V]{manhai_tac20} for a more detailed explanation on the switched-system case). }

This note considers families of impulsive systems (\ref{eq:nlsys-f}) that allow a global state weak linearization as per the following definition.
\begin{defin} \label{def:gswl}
  The family $\{\Sigma_\nu : \nu \in \Lambda\}$ of impulsive systems of the form~(\ref{eq:nlsys-f}) is said to admit a \emph{global state weak linearization} whenever there exist functions $A_\nu : \R_{\ge 0}\to \R^{n\times n}$ having locally bounded and Lebesgue measurable components, functions $\varphi_\nu : \R_{\ge 0} \times \R^n \times \R^m \to \R^n$, with $\varphi_{\nu}(t,\xi,\mu)$ Lebesgue measurable in $t$ and continuous in $(\xi,\mu)$, functions $R_\nu : \sigma_\nu \to \R^{n\times n}$ and $\psi_\nu : \sigma_\nu \times \R^n \times \R^m \to \R^n$, locally bounded and Lebesgue measurable functions $N^\nu:\R_{\ge 0}\to \R_{\ge 0}$, nonnegative constants $M$ and $c$, and\footnote{\color{blue}We write $\eta \in \K$ when $\eta : \R_{\ge 0} \to \R_{\ge 0}$ is continuous, strictly increasing and $\eta(0)=0$. We write $\eta\in\Ki$ when, in addition, $\eta$ is unbounded.} $\eta\in\K$, such that for all $\nu\in\Lambda$, $t\ge 0$, $\xi\in\R^n$ and $\mu\in\R^m$,
  \begin{subequations}
    \label{eq:maps-lin}
    \begin{align}
      \label{eq:fmap-lin}
      f_\nu(t,\xi,\mu) &= A_\nu(t) \xi + \varphi_\nu(t,\xi,\mu)\\
      \label{eq:jmap-lin}
      g_\nu(t,\xi,\mu) &= R_\nu(t) \xi + \psi_\nu(t,\xi,\mu)\\
      \label{eq:fpertBnd}
      |\varphi_\nu(t,\xi,\mu)| &\le N^\nu(t)|\xi|+(M|\xi| + c)\, \eta(|\mu|), & t &\notin \sigma_\nu,\\
      \label{eq:jpertBnd}
      |\psi_\nu(t,\xi,\mu)| &\le N^\nu(t) |\xi| + (M|\xi| + c)\, \eta(|\mu|), & t &\in \sigma_\nu.
    \end{align}
  \end{subequations}
\end{defin}
In~(\ref{eq:fpertBnd}), (\ref{eq:jpertBnd}) and from here on, $|\cdot|$ denotes Euclidean norm.
%
A family of systems admits a global state weak linearization whenever the flow and jump equations for every system can be written in a linear plus perturbation form, where the perturbation admits a specific type of bound that is affine on the state. Whenever a family of systems admits a global state weak linearization, for every $\nu\in\Lambda$, $t_0\ge 0$, $x_0\in \R^n$ and $u\in \U$, there exists a (not necessarily unique) forward-in-time solution $x:[t_0,T_x)\to \R^n$ of (\ref{eq:nlsys-f}) satisfying $x(t_0)=x_0$.
\begin{rem}
  The term $N^\nu(t)|\xi|$ in (\ref{eq:fpertBnd})--(\ref{eq:jpertBnd}) may depend on the specific system considered, i.e. may be different for each $\nu \in \Lambda$. The term $(M|\xi| + c)\eta(|\mu|)$ is not necessarily linear in $|\mu|$ and is the same for all the systems in the family. The bounds~(\ref{eq:fpertBnd})--(\ref{eq:jpertBnd}) reduce to $N^\nu(t)|\xi|$ under $\mu=0$. Note, however, that this does not mean that (\ref{eq:fmap-lin})--(\ref{eq:jmap-lin}) have to be linear in $\xi$ under $\mu=0$. \mer
\end{rem} 
\begin{rem}
  One important class of functions that satisfy bounds of the form (\ref{eq:fpertBnd}) or (\ref{eq:jpertBnd}) are those 
  that are zero at $(\xi,\mu)=(0,0)$ and Lipschitz in $(\xi,\mu)$. Another class is that of functions that can be written in the form $\sum_{j=1}^m \mu_j B_j(t) \xi + b(t) \mu$, where $\mu_j$ is the $j$-th component of $\mu$ and $B_j(t)$, $b(t)$ are matrices of appropriate dimensions. Therefore, (time-varying) bilinear systems admit a global state weak linearization. \mer
\end{rem} 

\subsection{Stability}
This note considers  stability properties where convergence is ensured not only as time elapses but also as the number of occurring impulses increases (see \cite{manhai_tac20, haiman_auto20} for background). For an impulse-time sequence $\sigma\in\Gamma$, let $n_{(s,t]}^\sigma$ denote the number of impulse times lying in the interval $(s,t]$, that is
\begin{align*}
n_{(s,t]}^\sigma := \#\big(\sigma \cap (s,t]\big).
\end{align*}
The superscript $^\sigma$ is removed whenever the corresponding impulse-time sequence $\sigma$ is clear from the context. Given $u\in \U$, define
\begin{align*}
\|u\|_{\infty} &:=\sup_{t\ge 0}|u(t)|,\quad\text{and for a given $\rho\in \K$,}\\
\|u\|_{\sigma,\rho} &:= \int_0^{\infty} \rho(|u(s)|)ds+\sum_{s\in \sigma \cap (0,\infty)}\rho(|u(s)|).
\end{align*}
\begin{defin}
  \label{def:stability}
  A system $\Sigma$ of the form~(\ref{eq:nlsys}) is said to be
  \begin{enumerate}[a)]
  \item \label{item:S-0-UAS} Strongly Zero-input Uniformly Asymptotically Stable (S-0-UAS) if there exist $r>0$ and\footnote{\color{blue}We write $\beta\in\KL$ when $\beta:\R_{\ge 0} \times \R_{\ge 0} \to \R_{\ge 0}$ satisfies $\beta(\cdot,t) \in \K$ for every $t$, $\beta(r,\cdot)$ is decreasing and $\lim_{t\to\infty} \beta(r,t) = 0$ for every $r$.} $\beta\in\KL$ such that its solutions satisfy
    \begin{align}
      \label{eq:0guas-ineq}
      |x(t)| &\le \beta(|x(t_0)|,t-t_0+n_{(t_0,t]}),
    \end{align}
    for all $t\ge t_0\ge 0$ whenever $|x(t_0)|\le r$ and $u(t) \equiv 0$. 
  \item Strongly Zero-input Globally Uniformly Asymptotically Stable (S-0-GUAS) if there exists $\beta\in\KL$ such that every solution satisfies~(\ref{eq:0guas-ineq}) for all $t\ge t_0\ge 0$ whenever $u(t) \equiv 0$.
  \item Strongly Zero-input (Globally) Uniformly Exponentially Stable (S-0-(G)UES) if it is S-0-(G)UAS and there exists $\lambda > 0$ and $K\ge 1$ such that (\ref{eq:0guas-ineq}) is satisfied with $\beta(s,t) = Kse^{-\lambda t}$. 
  \item Strongly Input-to-State Stable (S-ISS) if there exist $\beta\in\KL$ and $\gamma\in\K$ such that for all $t\ge t_0\ge 0$ the solutions satisfy
    \begin{align} \label{eq:s-iss}
      |x(t)| &\le \beta(|x(t_0)|,t-t_0+n_{(t_0,t]}) + \gamma(\|u\|_{\infty}).
    \end{align}
  \item Strongly integral ISS (S-iISS) if there exist $\beta\in\KL$ and $\gamma, \rho\in\K$ such that for all $t\ge t_0\ge 0$ the solutions satisfy
    \begin{align}
      |x(t)| &\le \beta(|x(t_0)|,t-t_0+n_{(t_0,t]}) + \gamma(\|u\|_{\sigma,\rho}).
    \end{align}
  \item \label{item:S-ISS-si} S-ISS under small inputs (S-ISS s.i.) with input threshold $R>0$ if there exist $\beta\in\KL$ and $\gamma\in\K$ such that the solutions satisfy (\ref{eq:s-iss}) for all $t\ge t_0\ge 0$ whenever $\|u\|_{\infty}\le R$.
  \end{enumerate}
{\color{blue} The definitions of "weak'' properties \ref{item:S-0-UAS})--\ref{item:S-ISS-si}), are obtained replacing $\beta(|x_0|,t-t_0+n_{(t_0,t]})$ by $\beta(|x_0|,t-t_0)$. For referring to the weak properties we replace S by W in the acronyms.}
\end{defin}
\begin{defin}
  \label{def:stability-p}
  A family of systems $\{\Sigma_\nu : \nu \in \Lambda \}$ of the form~(\ref{eq:nlsys-f}) is said to satisfy any of the properties in Definition~\ref{def:stability} (strong or {\color{blue} weak}) whenever the corresponding bound on the solutions holds uniformly for every system in the family, i.e., whenever the bounding functions and constants are the same for every system in the family.
\end{defin}
\subsection{The transition matrix}
The solution of an impulsive system of the form
\begin{subequations}
  \label{eq:impsimp}
  \begin{align}
    \dot x(t) &= A_\nu(t) x(t) + v_\nu(t), &t &\notin \sigma_\nu\\
    x(t) &= R_\nu(t) x(t^-) + w_\nu(t), & t &\in \sigma_\nu,
  \end{align}
\end{subequations}
with $A_\nu: \R_{\ge 0} \to \R^{n\times n}$ and $v_\nu:\R_{\ge 0}\to \R^n$ locally bounded and Lebesgue measurable, can be written as
\begin{align}
  x(t) &= \Phi^\nu(t,t_0) x(t_0) + \int_{t_0}^t \Phi^\nu(t,s) v_\nu(s) ds\ +\notag\\
  \label{eq:impsimpsol}
  &\hspace{5mm} + \sum_{s\in \sigma\cap (t_0,t]} \Phi^\nu(t,s) w_\nu(s),
\end{align}
where the transition matrix $\Phi^\nu(t,s)$, for $0 \le s \le t$, is defined as
\begin{align*}
&\scriptstyle    \Phi_c^\nu(t,s) & &\text{ if }\sigma_\nu \cap (s,t] = \emptyset,\\
& \scriptstyle  \Phi_c^\nu(r,r_k) R_k \cdots R_2 \Phi_c^\nu(r_2,r_1) R_1 \Phi_c^\nu(r_1,s) & &\text{ if }t\notin\sigma_\nu,\\
&\scriptstyle R_k \Phi_c^\nu(r_k,r_{k-1}) \cdots R_2 \Phi_c^\nu(r_2,r_1) R_1 \Phi_c^\nu(r_1,s) & &\text{ if }t\in\sigma_\nu,
\end{align*}
where $\Phi_c^\nu(\cdot,\cdot)$ is the transition matrix of $\dot x=A_\nu(t)x$, $R_i := R_\nu(r_i)$, and $r_1<\cdots<r_k$ are the impulse times lying in $(s,t]$.

It can be shown that for a single system (\ref{eq:impsimp}) the properties S-0-UAS (W-0-UAS), S-0-GUAS (W-0-GUAS) and S-0-GUES (W-0-GUES) are equivalent. This equivalence also holds for a family of systems of the form~(\ref{eq:impsimp}) for all $\nu\in\Lambda$. In addition, this family is S-0-GUES or W-0-GUES if and only if there exist $K\ge 1$ and $\lambda > 0$ such that for all $\nu\in\Lambda$ and $0\le s \le t$, 
\begin{align}
  \label{eq:PhiGUESbnd}
  \text{S-0-GUES: } \|\Phi^\nu(t,s)\| &\le K \exp \left[-\lambda(t-s+n_{(s,t]}^{\sigma_\nu})\right],\\
  \text{W-0-GUES: }  \|\Phi^\nu(t,s)\| &\le K \exp \left[-\lambda(t-s)\right],
   \label{eq:PhistGUESbnd}
\end{align}
{\color{blue} where $\|B\|=\max_{|\xi|=1}|B\xi|$ is the induced norm of the matrix $B$.}
\section{Main Results}

The main result is given in Section~\ref{sec:global-strong-iss}. This gives conditions for S-iISS, S-ISS and S-ISS s.i. of a family of impulsive systems that admits a global state weak linearization. Section~\ref{sec:standard-iss} first shows that the results do not remain valid if stability is not of the strong type but that the results can be recovered for {\color{blue}weak} stability by imposing constraints on the impulse-time sequence. Section~\ref{sec:linear-local} provides a result for establishing local exponential stability based on a first-order approximation. The results in Sections~\ref{sec:standard-iss} and~\ref{sec:linear-local} follow as corollaries to the main result.

\subsection{Global Strong (i)ISS}
\label{sec:global-strong-iss}

Theorem~\ref{thm:main} shows that a family of systems that admit a strongly stable global state weak linearization is both S-ISS under small inputs and S-iISS, provided the zero-input (linear) term of the perturbation bounds is appropriately small. 
\begin{teo}
  \label{thm:main}
  Consider a family $\{\Sigma_\nu : \nu \in \Lambda \}$ of impulsive systems of the form (\ref{eq:nlsys-f}). Suppose that this family admits a global state weak linearization so that (\ref{eq:maps-lin}) holds, and let $\Phi^\nu$, for every $\nu\in\Lambda$, denote the transition matrix of the associated system (\ref{eq:impsimp}). Suppose that
  there exist $K\ge 1$ and $\lambda > 0$ such that (\ref{eq:PhiGUESbnd}) holds for all $\nu\in\Lambda$ and $0 \le s \le t$. Suppose that there exists $\bar N<\frac{\lambda}{Ke^{\lambda}}$ such that $\theta^\nu(t) := \max\{N^\nu(t)-\bar N,0\}$ satisfies $\sup_{\nu\in\Lambda} \left( \int_0^{\infty} \theta^\nu(s)ds + \sum_{t\in\sigma_\nu} \theta^\nu(t)\right) <\infty$. Then, the family $\{\Sigma_\nu : \nu \in \Lambda \}$ is:
\begin{enumerate}[a)]
\item \label{item:S-0-GUES-iISS} S-0-GUES and S-iISS.
\item \label{item:ISSsi} S-ISS s.i. with input threshold $R < \eta^{-1}(\frac{\lambda - \bar{N}Ke^\lambda}{KMe^{\lambda}})$ if $M>0$.
\item \label{item:ISS2} S-ISS if $M=0$.
\end{enumerate}
In \ref{item:ISSsi})--\ref{item:ISS2}) one may take $\beta(s,t) = \bar K s e^{-\bar\lambda t}$ and $\gamma(s) = L\eta(s)$ for the bound (\ref{eq:s-iss}), for some $\bar K\ge 1$ and $\bar\lambda, L > 0$.
\end{teo}
{\color{blue}The constant $\bar N$ and the function $\theta^\nu$ in the statement of Theorem~\ref{thm:main} can be interpreted as the constant and vanishing parts, respectively, of the zero-input perturbation bounding coefficient $N^\nu(\cdot)$ in (\ref{eq:fpertBnd})--(\ref{eq:jpertBnd}). For stability to be possible, the constant part $\bar N$ should not be too large. The condition on the vanishing part $\theta^\nu(\cdot)$ actually ensures that stability is uniform over all systems in the family. The consideration of the constant and vanishing parts separately allows to avoid overly conservative requirements on $N^\nu(\cdot)$.
}

The proof of Theorem \ref{thm:main} requires the following lemmas.
\begin{lema}
  \label{lem:aux}
  Let $g,x: [a,b] \to \R^n$, $N,h: [a,b]\to \R_{\ge 0}$, and
  \begin{align} 
    |g(t)|\le N(t)|x(t)|+h(t)\quad \forall t\in [a,b].
  \end{align}
  Then, there exists $B: [a,b] \to \R^{n\times n}$ such that $\|B(t)\|\le N(t)$ and $|g(t)-B(t)x(t)|\le h(t)$ for all $t\in [a,b]$. If $g$, $z$, $N$ and $h$ are Lebesgue measurable, then the existence of a Lebesgue measurable $B(\cdot)$ can be ensured.
\end{lema}
\begin{lema}
  \label{lem:tmp}
  Consider the family of linear impulsive systems
  \begin{subequations}
    \label{eq:impsimpp}
    \begin{align}
      \dot x(t) &= [A_\nu(t)+B_\nu(t)] x(t) + \tilde v_\nu(t), &t &\notin \sigma_\nu,\\
      x(t) &= [R_\nu(t)+B_\nu^*(t)] x(t^-) + \tilde w_\nu(t), & t &\in \sigma_\nu,
    \end{align}
  \end{subequations}
  with $\sigma_\nu = \{\tau_k^\nu\}_{k=0}^\infty \in \Gamma$. Let $\Phi^\nu$ be the transition matrix of the associated system~(\ref{eq:impsimp}) and suppose that there exist $K\ge 1$ and $\lambda > 0$ such that (\ref{eq:PhiGUESbnd}) holds for all $\nu\in\Lambda$ and $0\le s \le t$. Let $B_\nu(\cdot)$ be Lebesgue measurable and $B_\nu^*(\cdot)$ such that there exists $\bar N < \frac{\lambda}{Ke^{\lambda}}$ so that $\theta_c^\nu(t) := \max\{\|B_\nu(t)\|-\bar N, 0 \}$ satisfies $\Theta_c := \sup_{\nu\in\Lambda} \int_0^{\infty}\theta_c^\nu(s)ds<\infty$ and $\theta_d^\nu(t) := \max\{ \|B_\nu^*(t)\|-\bar N, 0 \}$ satisfies $\Theta_d := \sup_{\nu\in\Lambda}\left(\sum_{t\in\sigma_\nu} \theta_d^\nu(t) \right) < \infty$. Then, the transition matrix $\Phi_p^\nu$ of (\ref{eq:impsimpp}) satisfies
  \begin{align}
    \label{eq:tmpbound}
    &\|\Phi_p^\nu(t,s)\|\le \hat K e^{-\hat\lambda \left( t - s + n_{(s,t]}^{\sigma_\nu} \right)} \quad \forall \nu\in\Lambda, \ \forall \:0\le s\le t,\\
     &\hat \lambda :=\lambda-Ke^\lambda \bar N,\quad \hat K := K \exp\left(\displaystyle K \left[ \Theta_c + e^{\lambda} \Theta_d \right] \right). 
  \end{align}
\end{lema}
\begin{IEEEproof}[Proof of Theorem \ref{thm:main}] \ref{item:S-0-GUES-iISS}) Let $x:[t_0,T_x)\to \R^n$, with $t_0\ge 0$, be a maximally defined solution of (\ref{eq:nlsys-f}) corresponding to some $\nu\in\Lambda$, $\sigma_\nu \in \Gamma$,
    and an input $u\in \U$ with $\|u\|_{\sigma_\nu,\eta}<\infty$, where $\eta \in \K$ is the function appearing in (\ref{eq:fpertBnd})--(\ref{eq:jpertBnd}). Let $x^*(t_0)=x_0$ and $x^*(t)=x(t^-)$ for all $t\in (t_0,T_x)$. Let $t_0<T<T_x$ and $J=[t_0,T]$. Define $g,g^* : J\to \R^n$ via $g(t)=\varphi_\nu(t,x(t),u(t))$ and $g^*(t)=\psi_\nu(t,x^*(t),u(t))$ if $t\in\sigma_\nu$ and $g^*(t) = 0$ otherwise, and $h,h^*:J\to \R_{\ge 0}$ via $h(t)=(M|x(t)|+c)\eta(|u(t)|)$ and $h^*(t)=(M|x^*(t)|+c)\eta(|u(t)|)$. Note that $g$ and $h$ are Lebesgue measurable, and $|g(t)|\le N^\nu(t)|x(t)|+h(t)$ and $|g^*(t)|\le N^\nu(t) |x^*(t)|+h^*(t)$. By applying Lemma~\ref{lem:aux} there exist $B_\nu : J \to \R^{n\times n}$ Lebesgue measurable such that for all $t\in J$, $\|B_\nu(t)\| \le N^\nu(t)$ and $|g(t)-B_\nu(t)x(t)|\le h(t)$, and also $B_\nu^* : J \to \R^{n\times n}$ with $\|B_\nu^*(t)\| \le N^\nu(t)$ and $|g^*(t)-B_\nu^*(t)x(t^-)|\le h^*(t)$.

Extend $B_\nu$ and $B_\nu^*$ as $0$ outside their domains. Then $x$ restricted to $J$ is a solution of (\ref{eq:impsimpp}) with $\tilde v_\nu(t)=g(t)-B_\nu(t)x(t)$ and $\tilde w_\nu(t)=g^*(t)-B_\nu^*(t)x^*(t)$. Therefore, $x(t)$ can be written as in (\ref{eq:impsimpsol}), with $\Phi_p^\nu$ instead of $\Phi^\nu$. The assumptions on $\bar N$ and the facts that $\|B_\nu(t)\| \le N^\nu(t)$ and $\|B^*_\nu(t)\| \le N^\nu(t)$ cause $B_\nu$ and $B^*_\nu$ to satisfy the hypotheses of Lemma \ref{lem:tmp}. Therefore, (\ref{eq:tmpbound}) holds. Since, in addition, $|\tilde v_\nu(t)|\le h(t)$ and $|\tilde w_\nu(t)|\le h^*(t)$ for all suitable $t\in J$ and from the definitions of $h$  $h^*$, it follows that for all $t\in J$
 \begin{align}
    |x(t)| &\le \hat Ke^{-\hat \lambda(t-t_0+n_{(t_0,t]})} |x(t_0)| \notag \\
    &\quad + \hat K \int_{t_0}^t e^{-\hat \lambda(t-s+n_{(s,t]})} (M|x(s)|+c) \eta(|u(s)|) ds \label{eq:xbndcomp}\\
    &\quad + \hat K \sum_{s\in \sigma_\nu\cap (t_0,t]} e^{-\hat \lambda(t-s+n_{(s,t]})} (M|x(s^-)|+c) \eta(|u(s)|)\notag
  \end{align}
 Define $\gamma(a,b)=b-a+n_{(a,b]}$, $0\le a<b$, $y(s)=|x(s)|e^{\hat \lambda \gamma(t_0,s)}$, $t_0<s$, and multiply the latter inequality by $e^{\hat \lambda \gamma(t_0,t)}$. By using 
the facts that $n_{(t_0,t]} - n_{(s,t]} = n_{(t_0,s]}$ for every $t_0 \le s \le t$ and that whenever $s\in\sigma_\nu \cap (t_0,t]$, 
 \begin{align*}
    e^{\hat \lambda} y(s^-) = |x(s^-)| e^{\hat \lambda(s-t_0+n_{(t_0,s)}+1)} = |x(s^-)| e^{\hat \lambda \gamma(t_0,s)}, 
 \end{align*}
 we have that for all $t\in J$
 \begin{align*}
   y(t) &\le p(t)+\int_{t_0}^t a(s) y(s) +\sum_{s\in \sigma_{\nu}\cap(t_0,t]} b(s)y(s^-),\\
     \text{with}\quad
     p(t) &:=\hat K  \left [ |x(t_0)| + c \left (\int_{t_0}^t e^{\hat \lambda \gamma(t_0,s)} \eta(|u(s)|) ds \right . \right . \notag \\
       &\makebox[.7in]{} \left .\left .+ \sum_{s\in \sigma_{\nu}\cap(t_0,t]} e^{\hat \lambda \gamma(t_0,s)} \eta(|u(s)|) \right ) \right ],\\
       a(t) &:=\hat K M \eta(|u(t)|)\quad\makebox{and}\quad b(t):= e^{\hat \lambda} \hat K M \eta(|u(t)|).
\end{align*}
Taking into account that $p(\cdot)$ is nondecreasing and applying Gronwall inequality (e.g. \cite[Lemma~3]{simbai_jmaa85}), we get
\begin{align*}
y(t)\le p(t) e^{\int_{t_0}^t a(s) ds} \prod_{s\in \sigma_{\nu}\cap(t_0,t]} (1+b(s))
\end{align*}
Since $\prod_{s\in \sigma_{\nu}\cap(t_0,t]} (1+b(s))\le e^{\sum_{s\in \sigma_{\nu}\cap(t_0,t]}b(s)}$, because $1+b\le e^b$ for all $b\ge 0$, and $\int_{t_0}^t a(s) ds + \sum_{s\in \sigma_{\nu}\cap(t_0,t]}b(s)\le e^{\hat \lambda} \hat K M\|u\|_{\sigma_{\nu},\eta}$, defining $\kappa=e^{\hat \lambda} \hat K M$, it follows that 
\begin{align*}
y(t)\le p(t) e^{\kappa \|u\|_{\sigma_{\nu},\eta}}.
\end{align*}
By multiplying the latter inequality by $\hat Ke^{-\hat \lambda\gamma(t_0,t))}$ we obtain
\begin{align*}
|x(t)|\le p(t) e^{-\hat \lambda\gamma(t_0,t)} e^{\kappa \|u\|_{\sigma_{\nu},\eta}}.
\end{align*}
Since 
\begin{align*}
p(t) e^{-\hat \lambda(t_0,t)} \le \hat K e^{-\hat \lambda \gamma(t_0,t)}|x_0|+c\hat K \|u\|_{\sigma_{\nu},\eta},
\end{align*}
$|x(t)|\le (\hat K e^{-\hat \lambda \gamma(t_0,t)}|x_0|+c\hat K \|u\|_{\sigma_{\nu},\eta}) e^{\kappa \|u\|_{\sigma_{\nu},\eta}}$.

Let $\alpha(r):=\ln(1+r)$. Then $\alpha \in \Ki$. By applying $\alpha$ in both sides of the latter inequality, and using that $\ln(1+a)\le a$ and $\ln(1+ae^b)\le \ln((1+a)e^b)\le a+b$ for all $a,b\ge 0$, 
\begin{align*}
\alpha(|x(t)|)\le \hat K e^{-\hat \lambda\gamma(t_0,t)}|x_0|+c\hat K \|u\|_{\sigma_{\nu},\eta}+ {\kappa \|u\|_{\sigma_{\nu},\eta}}.
\end{align*}
By applying $\zeta=\alpha^{-1}\in \Ki$ in both sides of the inequality, and using that $\zeta(a+b)\le \zeta(2a)+\zeta(2b)$ for all $a,b\ge 0$, it follows that
\begin{align*}
|x(t)|\le \zeta(2\hat K e^{-\hat \lambda \gamma(t_0,t)}|x_0|)+\zeta(2(c\hat K + \kappa) \|u\|_{\sigma_{\nu},\eta} ).
\end{align*}
Let $\beta(r,s)=\zeta(2\hat K e^{-\hat \lambda s} r)$ and $\rho(r)=\zeta(2(c\hat K+\kappa) r)$. Then $\beta \in \KL$, $\rho \in \Ki$ and, for all $t\in J=[t_0,T]$,
\begin{align}
|x(t)|\le \beta(|x_0|,t-t_0+n_{(t_0,t]})+\rho(\|u\|_{\sigma_{\nu},\eta}).
\end{align}
Since $T$ is any number satisfying $t_0<T<T_x$, it follows that $x$ is bounded on $[t_0,T_x)$ and therefore $T_x=\infty$ and the family is S-iISS. 

Let $x$ and $J$ be as above. If $u=0$, then $h=h^*=0$ and therefore $\tilde v_\nu=0$ and $\tilde w_\nu=0$. Then, from (\ref{eq:xbndcomp}) it follows that
\begin{align*}
|x(t)|\le \hat K e^{-\hat \lambda (t-s+n_{(t_0,t]})}|x(t_0)| 
\end{align*}
for all $t_0\le t\le T$ and all $T>t_0$ (since $T_x=\infty$). Therefore, the family $\{\Sigma_\nu : \nu \in \Lambda\}$ is S-0-GUES and item~\ref{item:S-0-GUES-iISS}) follows.
  
By causality, S-iISS implies that maximally defined solutions corresponding to any initial time $t_0$ and input $u \in \U$ are defined for all $t\ge 0$. This is shown as follows. If $x:[t_0,T_x)\to \R^n$ is a maximally defined solution corresponding to $u\in \U$ and $T_x<\infty$, then $|x(t)|\to \infty $ as $t\to T_x^-$. If we define $\bar u(t)=u(t)$ for $t<T_x$ and $\bar u(t)=0$ otherwise, then $\|\bar u\|_{\sigma_\nu,\eta}<\infty$ and $x$ restricted to $[t_0,T_x)$ is a solution corresponding to $\bar u$. Then the S-iISS property implies that $x$ is bounded on $[t_0,T_x)$ arriving to a contradiction.

  \ref{item:ISSsi}) Since $\bar N<\frac{\lambda}{Ke^{\lambda}}$, then $R>0$ satisfying $R<\eta^{-1}(\frac{\lambda - \bar{N}Ke^\lambda}{KMe^{\lambda}})$ exists. Then, $M\eta(R)<\frac{\lambda}{Ke^{\lambda}}-\bar N$. Let $x : [t_0,\infty)\to \R^n$ be a solution of (\ref{eq:nlsys-f}) corresponding to some $\nu\in\Lambda$ and an input $u$ with $\|u\|_{\infty}\le R$. Let $T>t_0$ and let $J$, $x^*$, $g$ and $g^*$ be as above. Define $\check N(t)=\bar N+M\eta(R)+\theta^\nu(t)$ and $\bar h: J\to \R_{\ge 0}$ via $\bar h(t)=c\eta(|u(t)|)$. Note that  $|g(t)|\le \check N(t)|x(t)|+\bar h(t)$ and $|g^*(t)|\le \check N(t)|x(t^-)|+\bar h(t)$ for all $t\in J$. By applying Lemma \ref{lem:aux} again, there exist $\bar B, {\bar B}^*:J\to \R^{n\times n}$, with $\bar B(\cdot)$ Lebesgue measurable, such that for all $t\in J$, $\max\{\|\bar B(t)\|,\|{\bar B}^*(t)\|\}\le \check N(t)$, $|g(t)-\bar B(t)x(t)|\le \bar h(t)$, $|g^*(t)-{\bar B}^*(t)x^*(t)|\le \bar h(t)$. Extend $\bar B$ and ${\bar B}^*$ to $[0,\infty)$ by defining as 0 outside $J$. Then $x$ restricted to $J$ is a solution of (\ref{eq:impsimpp}) with, respectively, $\bar B$, ${\bar B}^*$ instead of $B_\nu$,$B_\nu^*$, and with $\tilde v_\nu(t) = g(t)-\bar B(t)x(t)$ and $\tilde w_\nu(t) = g^*(t)-B^*(t)x(t^-)$. Since $\bar N+M\eta(R)<\frac{\lambda}{Ke^{\lambda}}$, by Lemma~\ref{lem:tmp} the corresponding transition matrix $\bar \Phi_p^\nu$ satisfies 
  \begin{align}
    \label{eq:tmpboundbis}
    \|\bar \Phi_p^\nu(t,s)\| &\le \bar K e^{-\bar \lambda (t-s+n_{(s,t]}^{\sigma_\nu})}\quad \forall \:0\le s\le t,
  \end{align}
  with $\bar \lambda =\lambda-Ke^\lambda (\bar N+M\eta(R))>0$ and $\bar K =K e^{K\Theta_c + e^{\lambda} \Theta_d}$. Writing $x(t)$ as in (\ref{eq:impsimpsol}), with $\bar \Phi_p^\nu$, $\tilde v_\nu$ and $\tilde w_\nu$ instead of, respectively, $\Phi^\nu$, $v_\nu$ and $w_\nu$, applying the norm and using its properties, it follows that for all $t\in J$
 \begin{align}
   |x(t)| &\le \bar Ke^{-\bar \lambda \gamma(t_0,t)} |x(t_0)|
  + \bar K c \int_{t_0}^t e^{-\bar \lambda \gamma(s,t)} \eta(|u(s)|) ds\notag\\
    &\quad + \bar K c \sum_{s\in \sigma_\nu\cap (t_0,t]} e^{-\bar \lambda \gamma(s,t)} \eta(|u(s)|) \notag \\
&\le \bar K e^{-\bar \lambda \gamma(t_0,t)} |x(t_0)|\notag \\
  &\quad + \bar K c\eta(\|u\|_{\infty}) \left( \int_{t_0}^t e^{-\hat \lambda(t-s)} ds + \sum_{s\in \sigma_\nu\cap (t_0,t]} e^{-\bar\lambda n_{(s,t]}}\right) \notag \\
        &\le \bar K e^{-\bar \lambda \gamma(t_0,t)} |x(t_0)|+\bar K c\left(\frac{1}{\bar \lambda}+\frac{1}{1-e^{-\bar \lambda}}\right)\eta(\|u\|_{\infty}) \label{eq:bound3}
      \end{align}
where we have used that for all $r,n\ge 0$, then $e^{-\bar\lambda (r+n)} \le \min\{e^{-\bar\lambda r},e^{-\bar\lambda n}\}$ and
\begin{align*}
  \int_{t_0}^t e^{-\bar \lambda(t-s)}ds < \frac{1}{\bar \lambda}
  \quad\text{and }
  \sum_{s\in \sigma_\nu\cap (t_0,t]} e^{-\bar \lambda n_{(s,t]}}<\frac{1}{1-e^{-\bar \lambda}}
\end{align*}
Since $T$ is arbitrary, then (\ref{eq:bound3}) holds for all $t\ge t_0$. Hence, for all $u$ such that $\|u\|_{\infty}\le R$, the corresponding solution satisfies
\begin{align*}
  |x(t)|\le \beta(|x(t_0)|,t-t_0+n_{(t_0,t]}^{\sigma_\nu})+\rho(\|u\|_{\infty})
\end{align*}
for all $t\ge t_0$, with the function $\beta \in \KL$, $\beta(s,t)=\bar K s e^{-\bar \lambda t}$ and the function $\rho \in \K$ given by $\rho(s)=\bar K c\left(\dfrac{1}{\bar \lambda}+\dfrac{1}{1-e^{-\bar \lambda}}\right)\eta(s)$. Since neither $\beta$ nor $\rho$ depend on $\nu\in\Lambda$, then the family $\{\Sigma_\nu : \nu\in\Lambda\}$ is S-ISS s.i.

\ref{item:ISS2}) The proof for the case $M=0$ is obtained by simplifying the above derivations and is thus straightforward.
\end{IEEEproof}
The proof of Theorem~\ref{thm:main} is original in that each solution of an impulsive system that admits a global state weak linearization is written as the solution to a linear system of the form~(\ref{eq:impsimpp}) where the matrices $B(t)$ and $B^*(t)$ are constructed based on the perturbations $\varphi_\nu,\psi_\nu$, their bounds, and the specific solution $x(\cdot)$ considered.

\subsection{\color{blue}Weak ISS}
\label{sec:standard-iss}

The results of Theorem~\ref{thm:main} do not remain valid if the linear system (\ref{eq:impsimp}) is assumed to be only {\color{blue} W-0-GUES} instead of S-0-GUES.
\begin{example}
  \label{ex:1}
  Consider the linear scalar impulsive system
  \begin{align}
    \dot{x}(t) &=-x(t), \; t\notin \sigma, &
    x(t) &= x(t^-), \; t\in \sigma,
  \end{align}
with $\sigma=\{\tau_k\}_{k=1}^\infty$, defined recursively by $\tau_1=1$, $\tau_{k+1}=\tau_k+1/k$. For every initial state $x_0\in \R$ and initial time $t_0\ge 0$, the corresponding solution is $x(t)=x_0 e^{-(t-t_0)}$. This system is thus {\color{blue} W-0-GUES}. For $\delta>0$, consider the system obtained by slightly perturbing the jump equation so that
\begin{align}
  \label{eq:psys}
  \dot{x}(t) &=-x(t), \; t\notin \sigma, &
  x(t) &=x(t^-)+\delta x(t^-), \; t\in \sigma.
\end{align}
The system~(\ref{eq:psys}) clearly admits a global state weak linearization satisfying the bounds (\ref{eq:fpertBnd}) and (\ref{eq:jpertBnd}) with $N^\nu(t)\equiv \delta$, $M=0$, $c=0$ and arbitrary $\eta \in \K$. If $x(t)$ is a solution of~(\ref{eq:psys}), then for every $\tau_k>t_0$, $x(\tau_{k+1})=(1+\delta)e^{-\Delta_k}x(\tau_k)$, where $\Delta_k=\tau_{k+1}-\tau_k$. Since $\Delta_k = 1/k$, there exists $d>1$ such that $(1+\delta)e^{-\Delta_k}>d$ for all $k$ large enough. In consequence, $x(\tau_k)\to \infty$ whenever $x_0 \neq 0$ and (\ref{eq:psys}) is not {\color{blue} W-0-GUES}.\mer
 \end{example}

Example \ref{ex:1} suggests that some condition on the distance between consecutive impulse times is needed for the validity of a result analogous to Theorem \ref{thm:main} when just {\color{blue} W-0-GUES} is assumed. A sufficient condition for this purpose is that $\sigma_{\nu}\in \Gamma[N_0,\tau_D]$, with $\Gamma[N_0,\tau_D]$ the set of sequences having chatter bound $N_0\in\N$ and average dwell-time $\tau_D>0$. More precisely, $\sigma \in \Gamma[N_0,\tau_D]$ if $n^{\sigma}_{(t_0,t]}\le N_0+\frac{t-t_0}{\tau_D}$ for all $0\le t_0<t$.
\begin{cor}
  \label{cor:main}
  Consider a family $\{\Sigma_\nu : \nu \in \Lambda \}$ of impulsive systems of the form (\ref{eq:nlsys-f}). Suppose that this family admits a global state weak linearization so that (\ref{eq:maps-lin}) holds, and let $\Phi^\nu$, for every $\nu\in\Lambda$, denote the transition matrix of the associated system (\ref{eq:impsimp}). Suppose that
  there exist $K\ge 1$ and $\lambda > 0$ such that (\ref{eq:PhistGUESbnd}) holds for all $\nu\in\Lambda$ and $0 \le s \le t$. Let $\{\sigma_{\nu}:\nu\in \Lambda\}\subset \Gamma[N_0, \tau_D]$ for some $N_0\in \N$ and $\tau_D>0$. Suppose that there exists $\bar N<\frac{\tilde \lambda}{\tilde Ke^{\tilde \lambda}}$, with $\tilde \lambda=\frac{\tau_D}{1+\tau_D} \lambda$ and $\tilde K=Ke^{N_0 \tilde \lambda}$, such that $\theta^\nu(t) := \max\{N^\nu(t)-\bar N,0\}$ satisfies $\sup_{\nu\in\Lambda} \left( \int_0^{\infty} \theta^\nu(s)ds + \sum_{t\in\sigma_\nu} \theta^\nu(t)\right) <\infty$. Then, the following hold for the family $\{\Sigma_\nu : \nu \in \Lambda \}$:
\begin{enumerate}[a)]
\item It is {\color{blue} W-0-GUES and W-iISS}.
\item It is {\color{blue} W-ISS if $M=0$ and W-ISS s.i. if $M>0$.}
\end{enumerate}
\end{cor}
\begin{IEEEproof}
  Let $\nu \in \Lambda$. By assumption, $n^{\sigma_\nu}_{(t_0,t]}\le N_0+\frac{t-t_0}{\tau_D}$ for all $0\le t_0<t$. Therefore, $t-t_0\ge \tau_D [n^{\sigma_\nu}_{(t_0,t]}-N_0]$, and then
\begin{align*}
\lambda (t-t_0)&= \left(1-\frac{1}{1+\tau_D}\right) \lambda(t-t_0) + \frac{\lambda}{1+\tau_D}(t-t_0)\\
&\ge \frac{\tau_D \lambda}{1+\tau_D}(t-t_0)+\frac{\lambda}{1 + \tau_D} \tau_D [n^{\sigma_\nu}_{(t_0,t]}-N_0]\\
&=\tilde \lambda (t-t_0+n^{\sigma_\nu}_{(t_0,t]})-\tilde \lambda N_0.
\end{align*}
In consequence, for every $0\le t_0 <t$ it follows that
\begin{align*}
\|\Phi^{\nu}(t,t_0)\| \le K e^{-\lambda (t-t_0)}
&\le K e^{\tilde \lambda N_0} e^{-\tilde \lambda (t-t_0+n^{\sigma_\nu}_{(t_0,t]})}.
\end{align*}
Since $\tilde K = K e^{\tilde \lambda N_0}$, we have proved that the family of linear systems (\ref{eq:impsimp}) satisfies (\ref{eq:PhiGUESbnd}) with $\tilde K$ and $\tilde \lambda$ instead of $K$ and $\lambda$. Applying Theorem~\ref{thm:main} and the fact that the strong stability properties imply the {\color{blue} weak ones}, Corollary~\ref{cor:main} is established.
\end{IEEEproof}
\begin{rem}
  The nontrivial part of Theorem 3.2 in \cite{liudou_scl14}, namely that (iv) implies (i), is a direct consequence of Corollary~\ref{cor:main}. The MiDT condition assumed in \cite[Theorem~3.2]{liudou_scl14} implies that the impulse-time sequence $\sigma$ belongs to $\Gamma[1,\Delta_{\inf}]$ and from the linear-plus-perturbation form of the system considered therein it is evident that the system admits a global state weak linearization.
    Moreover, Corollary~\ref{cor:main} implies that \cite[Theorem~3.2]{liudou_scl14} actually holds under weaker assumptions, specifically under $\sigma \in \Gamma[N_0,\tau_D]$. \mer
\end{rem}

\subsection{Linearization-based Local (Strong) Exponential Stability}
\label{sec:linear-local}

We next generalize, to the case of strong stability of impulsive systems, the standard result on local stability of nonlinear systems based on a first-order approximation. To keep matters simple, the result is given for a single system. The corresponding result for a family of systems will be true when the required bounds hold uniformly for every system in the family. Consider a system of the form~(\ref{eq:nlsys}) under zero input, with $f(t,0,0)=0$ for all $t\ge 0$, $g(t,0,0)=0$ for all $t\in \sigma$, $f(t,\xi,0)$ Lebesgue measurable in $t$ and $f(t,\cdot,0),g(t,\cdot,0)$ differentiable at $\xi = 0$ for every corresponding $t$, so that the matrices $A(t):=\frac{\partial f}{\partial \xi}(t,0,0)$ and $R(t):=\frac{\partial g}{\partial \xi}(t,0,0)$ exist. The corresponding linearized system under zero input is
  \begin{align}
  \label{eq:linearization}
    \dot{x}(t) &= A(t) x(t), \ t \notin\sigma, &
    x(t) &= R(t) x(t^-), \ t \in \sigma.
  \end{align}
\begin{cor} \label{cor:linearization}
  Suppose that the limit operations that define the derivatives for obtaining the matrices $A(t)$ and $R(t)$ are uniform over all suitable values of $t$, i.e. for every $\varepsilon > 0$ there exists $r>0$ such that whenever $|\xi| < r$, it happens that
  \begin{align}
    \label{eq:limf}
    |f(t,\xi,0)-A(t)\xi| &\le \varepsilon |\xi|,\quad\forall t\ge 0, \text{ and}\\
    \label{eq:limg}
    |g(t,\xi,0)-R(t)\xi| &\le \varepsilon |\xi|,\quad\forall t\in\sigma.
  \end{align}
Then, the following hold.
\begin{enumerate}[a)]
\item If the linearization (\ref{eq:linearization}) is S-0-GUES, then (\ref{eq:nlsys}) is S-0-UES.\label{item:lin-s0gues}
\item If the linearization (\ref{eq:linearization}) is {\color{blue} W-0-GUES} and $\sigma \in \Gamma[N_0,\tau_D]$ for some $N_0\in \N$ and $\tau_D>0$, then (\ref{eq:nlsys}) is {\color{blue} W-0-UES}.\label{item:lin-0gues}
\end{enumerate}
\end{cor}
\begin{IEEEproof} \ref{item:lin-s0gues}) Let $K\ge 1$ and $\lambda >0$ be the constants characterizing the S-0-GUES property of (\ref{eq:linearization}). Take $\varepsilon = \frac{\lambda}{2Ke^{\lambda}}$ and consider the corresponding $r>0$ according to the assumptions.
  Let $\Lambda = \{\nu\}$ be a single-element set. Let $\omega:\R^n\to [0,1]$ be a continuous function such that $\omega(\xi)=1$ if $|\xi|\le \frac{r}{2}$ and  $\omega(\xi)=0$ if $|\xi|\ge r$. Define $A_\nu \equiv A$, $R_\nu \equiv R$, $\varphi_\nu(t,\xi,\mu)=\omega(\xi) [f(t,\xi,0)- A(t)\xi]$ and $\psi_\nu(t,\xi,\mu)=\omega(\xi) [g(t,\xi,0) - R(t)\xi]$. Then (\ref{eq:fpertBnd})--(\ref{eq:jpertBnd}) hold with $N^\nu(t)=\bar N=\frac{\lambda}{2Ke^{\lambda}}$, $M=c=0$ and any $\eta\in \K$. It follows that the single-system family given by (\ref{eq:nlsys-f}), (\ref{eq:fmap-lin}) and (\ref{eq:jmap-lin}) satisfies the hypotheses of Theorem~\ref{thm:main} and is hence S-0-GUES. In consequence, there exist $\bar K, \bar \lambda>0$ such that the solutions with $u=0$ satisfy
$|x(t)|\le \bar K e^{\bar \lambda (t-t_0+n_{(t_0,t]})}|x(t_0)|$ for all $t\ge t_0$. The solutions of (\ref{eq:nlsys-f}) are solutions of (\ref{eq:nlsys}) under zero input (and viceversa) if $|x(t_0)|\le \dfrac{r}{2 \bar K}$, because they satisfy the condition $|x(t)|\le \frac{r}{2}$ for all $t\ge t_0$. Therefore, (\ref{eq:nlsys}) is S-0-UES.

\ref{item:lin-0gues}) Since (\ref{eq:linearization}) is {\color{blue} W-0-GUES} and  $\sigma \in \Gamma[N_0,\tau_D]$ then (\ref{eq:linearization}) is S-0-GUES (see the proof of Corollary \ref{cor:main}). By~\ref{item:lin-s0gues}), then (\ref{eq:nlsys}) is S-0-UES, and therefore {\color{blue} W-0-UES}.
\end{IEEEproof}
\begin{rem}
Corollary \ref{cor:linearization} improves the stability criterion based on first-order approximation in \cite[Theorem 37]{samper_book95}, since the latter assumes that the distance between consecutive impulse-times is bounded from below by a positive constant. Theorem 4.1 in \cite{yemic_tac98} also follows as a particular case. \mer
\end{rem}
{\color{blue}
  \section{Application to Switched Impulsive Systems}\label{sec:STVIS}
  
We next explain how the previous results can be applied to the stability analysis of switched time-varying impulsive systems. To keep matters simple, we consider switched impulsive systems where the switching instants coincide with impulse instants. However, the results can be applied to a more general class of systems (e.g. that in \cite[Section V]{manhai_tac20}).

\subsection{Casting switched systems as families of systems}
Consider switched impulsive systems defined as follows. Let  $\I =\{1,\ldots,\bar n\}$ be the set of switching modes. For $i\in \I$, the equation
  \begin{align}
    \label{eq:flow-map-i}
    \dot{x} &=\mathbf{f}_i(t,x,u),
    \end{align}
 is the dynamics of mode $i$ and $\mathbf{f}_i:\R_{\ge 0}\times\R^n \times \R^m \to \R^n$ is the corresponding flow map. For $i,j\in \I$ with $i\neq j$, $\mathbf{g}_{i,j}:\R_{\ge 0}\times \R^n \times \R^m\to \R^n$ denotes the reset map from mode $i$ to mode $j$.
 
Let $\nu:\R_{\ge 0}\to \I$ be a switching signal, {\em i.e.} $\nu$ is piecewise constant, right-continuous and with a finite number of discontinuities in each compact interval, and let $\sigma_{\nu}=\{t>0:\nu(t^-)\neq \nu(t)\}=\{\tau_1,\tau_2,\ldots\}$ be the set of (ordered) switching times of $\nu$. The switched system corresponding to the switching signal $\nu$ is the impulsive system 
\begin{subequations}
  \label{eq:swsys}
  \begin{align}
    \label{eq:swflow}
    \dot{x}(t) &= \mathbf{f}_{\nu(t)}(t,x(t),u(t)), & t &\notin\sigma_{\nu},\\
    \label{eq:swjump}
    x(t) &= \mathbf{g}_{\nu(t^-),\nu(t)}(t,x(t^-),u(t)), & t &\in \sigma_{\nu}.
  \end{align}
  \end{subequations}
Note that on each interval where mode $i$ is active (i.e. $\nu(t)=i$), the state $x$ is a solution of (\ref{eq:flow-map-i}) and when mode $i$ ends and mode $j$ begins (i.e $\nu(t^-)=i$ and $\nu(t)=j$), the state is reset to the value $\mathbf{g}_{i,j}(t,x(t^-),u(t))$.

Stability properties of~(\ref{eq:swsys}) which hold not only for a specific switching signal $\nu$ but also uniformly for all signals $\nu$ within some family $\Lambda$ (e.g. all signals having some dwell or average dwell time) can be analyzed by casting the set of switched systems corresponding to such switching signals as a family of time-varying impulsive systems, parametrized by the switching signal, as follows. For each switching signal $\nu$, let $\Sigma_{\nu}$ denote the switched system (\ref{eq:swsys}) corresponding to $\nu$ and note that it can be written in the form (\ref{eq:nlsys-f}) with 
\begin{align*}
  f_{\nu}(t,\xi,\mu) &:=\mathbf{f}_{\nu(t)}(t,\xi,\mu), &(t,\xi,\mu) &\in \R_{\ge 0} \times \R^n \times \R^m,\\ 
  g_{\nu}(t,\xi,\mu) &:=\mathbf{g}_{\nu(t^-),\nu(t)}(t,\xi,\mu), &(t,\xi,\mu) &\in \sigma_{\nu} \times \R^n \times \R^m. 
\end{align*}
In these expressions, $\nu$ denotes a function, $\nu\in\Lambda$, whereas $\nu(t)$ denotes the value of this function at time $t$, $\nu(t) \in \I$. 
With these definitions, the switched system (\ref{eq:swsys}) has some stability property which holds uniformly over $\Lambda$ if and only if the family $\{\Sigma_{\nu}:\nu \in \Lambda\}$ has the same stability property.

\subsection{Switched systems admitting a global state weak linerization}

Suppose that the maps $\mathbf{f}_i$ and $\mathbf{g}_{i,j}$ can be written in the form
\begin{align}
\mathbf{f}_i(t,\xi,\mu) &=\mathbf{A}_i(t)\xi+\hat{\varphi}_{i}(t,\xi,\mu) \label{eq:I}\\
\mathbf{g}_{i,j}(t,\xi,\mu) &=\mathbf{R}_{i,j}(t)\xi+\hat{\psi}_{i,j}(t,\xi,\mu) \label{eq:II}
\end{align}
with $\mathbf{A}_i:\R_{\ge 0}\to \R^{n\times n}$ with locally bounded and Lebesgue measurable entries, $\hat{\varphi}_{i}(t,\xi,\mu)$, Lebesgue measurable in $t$ and continuous in $(\xi,\mu)$ and $\mathbf{R}_{i,j}:\R_{\ge 0}\to \R^{n\times n}$. Suppose also that for all $t\ge 0$,
\begin{align}
      \label{eq:fpertBndsw}
      |\hat \varphi_i(t,\xi,\mu)| &\le \mathbf{N}^{i}(t)|\xi|+(M|\xi| + c)\, \eta(|\mu|), \\
      \label{eq:jpertBndsw}
      |\hat \psi_{i,j}(t,\xi,\mu)| &\le \mathbf{N}^{i,j}(t) |\xi| + (M|\xi| + c)\, \eta(|\mu|),
    \end{align}
with $N^{i},N^{i,j}:\R_{\ge 0}\to \R_{\ge 0}$, $N^i$ locally bounded and Lebesgue measurable, $M$ and $c$ nonnegative constants and $\eta \in \K$.
Then, it is straightforward that the family $\{\Sigma_{\nu}: \nu \in \Lambda\}$ admits a global state weak linearization as per Definition~\ref{def:gswl}, with
\begin{align*}
 A_{\nu}(t) &:=\mathbf{A}_{\nu(t)}(t), & R_{\nu}(t) &:=\mathbf{R}_{\nu(t^-),\nu(t)}(t),\\
\varphi_{\nu}(t,\xi,\mu) &:=\hat{\varphi}_{\nu(t)}(t,\xi,\mu), &
\psi_{\nu}(t,\xi,\mu) &:=\hat{\psi}_{\nu(t^-),\nu(t)}(t,\xi,\mu),\\
N^{\nu}(t) := &\mathbf{N}^{\nu(t)}(t),\;t\notin \sigma_{\nu}, &
N^{\nu}(t) := &\mathbf{N}^{\nu(t^-),\nu(t)}(t),\;t\in \sigma_{\nu},
 \end{align*}
 and the same constants $M$ and $c$.
This fact allows application of Theorem~\ref{thm:main} and Corollary~\ref{cor:main} to the analysis of switched systems.

As an example, consider a (nonlinear, time-varying) switched system~(\ref{eq:swsys}) and a family $\Lambda$ of switching signals $\nu$. Suppose the switched system admits a global state weak linearization where the linear part of the perturbation bounds is time-invariant, so that (\ref{eq:fpertBndsw})--(\ref{eq:jpertBndsw}) are satisfied with $\mathbf{N}^i(t)\equiv \bar N$ and $\mathbf{N}^{i,j}(t)\equiv \bar N$ (and therefore $N^{\nu}(t)\equiv \bar N$ for all $\nu$). 

In this setting, note that the transition matrix $\Phi^\nu(t,s)$ of the associated system~(\ref{eq:impsimp}) corresponding to a switching signal $\nu\in\Lambda$ is the transition matrix of the switched linear impulsive system $\dot x(t) = \mathbf{A}_{\nu(t)}(t) x(t)$ for $t\notin\sigma_\nu$, $x(t) = \mathbf{R}_{\nu(t^-),\nu(t)}(t) x(t^-)$ for $t\in\sigma_\nu$. The particularization of Theorem \ref{thm:main} and of Corollary~\ref{cor:main} to this case is as follows.
\begin{cor}
  \label{cor:switched}
  Suppose that there exist $K\ge 1$ and $\lambda > 0$ such that $\Phi^\nu(t,s)$ satisfies~(\ref{eq:PhiGUESbnd}) for all $\nu\in\Lambda$ and $0 \le s \le t$. If $\bar N<\frac{\lambda}{Ke^{\lambda}}$, then the considered switched system is:
\begin{enumerate}[a)]
\item S-0-GUES and S-iISS, uniformly over $\Lambda$.
\item S-ISS if $M=0$ and S-ISS s.i. if $M>0$, uniformly over $\Lambda$.
\end{enumerate}
\end{cor}
\begin{cor}
  \label{cor:wswitched}
  Suppose that there exist $K\ge 1$ and $\lambda > 0$ such that $\Phi^\nu(t,s)$ satisfies~(\ref{eq:PhistGUESbnd}) for all $\nu\in\Lambda$ and $0 \le s \le t$, and that for some $N_0\ge 1$ and $\tau_D>0$, $\sigma_{\nu}\in \Gamma[N_0,\tau_D]$ for all $\nu \in \Lambda$. If $\bar N<\frac{\tilde \lambda}{\tilde Ke^{\tilde \lambda}}$, with $\tilde \lambda=\frac{\tau_D}{1+\tau_D} \lambda$ and $\tilde K=Ke^{N_0 \tilde \lambda}$, then the considered switched system is:
\begin{enumerate}[a)]
\item W-0-GUES and W-iISS, uniformly over $\Lambda$.
\item W-ISS if $M=0$ and W-ISS s.i. if $M>0$, uniformly over $\Lambda$.
\end{enumerate}
\end{cor}

These results allow to establish stability of the original (nonlinear, time-varying) switched system 
by studying its linear part and the perturbation bound. 

}

\section{Proof of auxiliary lemmas}

\subsection*{Proof of Lemma~\ref{lem:aux}:}
  Let $t\in [a,b]$. If $g(t)=0$  or $x(t)=0$ define $B(t)=0$. In both cases $|g(t)-B(t)x(t)|=|g(t)|\le h(t)$. If $g(t)\neq 0$ and $x(t)\neq 0$, pick any orthogonal matrix $U$ such that $U\frac{x(t)}{|x(t)|}=\frac{g(t)}{|g(t)|}$ and define $B(t)=\min \left \{\frac{|g(t)|}{|x(t)|}, N(t) \right \}U$. If $\frac{|g(t)|}{|x(t)|}\le N(t)$, $B(t)x(t)=g(t)$ and then $|g(t)-B(t)x(t)|\le h(t)$. In case $\frac{|g(t)|}{|x(t)|}> N(t)$, $|g(t)-B(t)x(t)|=|g(t)|-N(t)|x(t)|\le h(t)$. Thus, the existence of a function $B: [a,b]\to \R^{n\times n}$ such that  $\|B(t)\|\le N(t)$ and $|g(t)-B(t)x(t)|\le h(t)$ for all $t\in [a,b]$ is established.

Next, let $g$, $x$, $h$ and $N$ be Lebesgue measurable. Define the function $F:[a,b]\times \R^{n\times n}\times \R\to \R$, via $F(t,B,y)=|g(t)-Bx(t)|-y$, and the set-valued map $\Omega:[a,b]\rightrightarrows \R^{n\times n}\times \R$, by $\Omega(t)=B_{N(t)}\times [0,h(t)]$, where, for $r\ge 0$, $B_r=\{A\in \R^{n\times n}:\|A\|\le r\}$. Then $F$ is measurable in $t$ and continuous in $(B,y)$, and $\Omega$ takes compact values and is measurable. Note that the function $B(t)$ defined above together with the function $y(t)=|g(t)-B(t)x(t)|$, $t\in [a,b]$, satisfy for all $t\in [a,b]$ the conditions $F(t,B(t),y(t))=0$ and $(B(t),y(t))\in \Omega(t)$. So, $0\in F(t,\Omega(t))$ for all $t\in [a,b]$.  Application of Theorem 7.1 in \cite{himmelberg_fm75} establishes the existence of a Lebesgue measurable function $(B^*,y^*):[a,b]\to \Omega(t)$ such that $F(t,B^*(t),y^*(t))=0$ for all $[a,b]$. Thus $\|B^*(t)\|\le N(t)$ and $|g(t)-B^*(t)x(t)|=y^*(t)\le h(t)$ for all $t\in [a,b]$.

\subsection*{Proof of Lemma~\ref{lem:tmp}:}
  Every solution of~(\ref{eq:impsimpp}) corresponding to $\tilde v_\nu \equiv 0$ and $\tilde w_\nu \equiv 0$ satisfies~(\ref{eq:impsimpsol}) with $v_\nu(t) = B_\nu(t) x(t)$ and $w_\nu(t) = B^*_\nu(t) x(t^-)$.
  Taking the norm and using its properties,
  \begin{align*}
    |x(t)| &\le \|\Phi^\nu(t,t_0)\| |x(t_0)| + \int_{t_0}^t \|\Phi^\nu(t,s)\| |B_\nu(s) x(s)| ds \\
    &\ + \sum_{s\in \sigma_\nu \cap (t_0,t]} \|\Phi^\nu(t,s)\| |B_\nu^*(s) x(s^-)|.
  \end{align*}
  Using~(\ref{eq:PhiGUESbnd}), then
  \begin{align*}
    |x(t)| &\le Ke^{-\lambda(t-t_0+n_{(t_0,t]}^{\sigma_\nu})} |x(t_0)|\\
      &\ + \int_{t_0}^t Ke^{-\lambda(t-s+n_{(s,t]}^{\sigma_\nu})} \|B_\nu(s)\| |x(s)| ds \\
    &\ + \sum_{s\in \sigma_\nu \cap (t_0,t]} Ke^{-\lambda(t-s+n_{(s,t]}^{\sigma_\nu})} \|B_\nu^*(s)\| |x(s^-)|.
  \end{align*}
  Define $y(t) := |x(t)| e^{\lambda(t-t_0+n_{(t_0,t]}^{\sigma_\nu})}$ and multiply the latter inequality by $e^{\lambda(t-t_0+n_{(t_0,t]}^{\sigma_\nu})}$. This gives
  \begin{align*}
    \scriptstyle  y(t) \le K|x(t_0)| + K \big(\int_{t_0}^t \|B_\nu(s)\| y(s) ds + \sum_{s\in \sigma_\nu \cap (t_0,t]} e^{\lambda} \|B_\nu^*(s)\| y(s^-) \big),
  \end{align*}
  using the facts that $n_{(t_0,t]} - n_{(s,t]} = n_{(t_0,s]}$ for every $t_0 \le s \le t$ and that whenever $s\in\sigma_\nu \cap (t_0,t]$, 
  \begin{align*}
    e^\lambda y(s^-) = |x(s^-)| e^{\lambda(s-t_0+n_{(t_0,s)}^{\sigma_\nu}+1)} = |x(s^-)| e^{\lambda(s-t_0+n_{(t_0,s]}^{\sigma_\nu})} . 
  \end{align*}
  Applying Gronwall inequality (e.g. \cite[Lemma~3]{simbai_jmaa85}) yields
  \begin{align}
    y(t) &\le K|x(t_0)| e^{K\int_{t_0}^t\|B_\nu(s)\| ds } \prod_{s\in \sigma_\nu\cap (t_0,t]}\hspace{-2mm}(1+K e^{\lambda} \|B_\nu^*(s)\|)  \notag\\
      &\le K|x(t_0)| \exp\left\{ {\scriptstyle
        K \left( \bar N(t-t_0) + \Theta_c + e^{\lambda} (\bar N n_{(t_0,t]}^{\sigma_\nu} + \Theta_d) \right)} \right\},\notag\\
        &\le \hat K|x(t_0)| e^{K\bar N e^\lambda (t-t_0 + n_{(t_0,t]}^{\sigma_\nu})} \label{eq:fin}
  \end{align}
where the assumptions on $\|B_\nu(t)\|$ and $\|B_\nu^*(t)\|$ and the fact that $1+b\le e^b$ for nonnegative $b$ have been employed.
 Multiplying (\ref{eq:fin}) by $e^{-\lambda(t-t_0+n_{(t_0,t]})}$ and recalling the definition of $y$ and $\hat\lambda$, yields
  \begin{align}
    \label{eq:pertlinsolexp}
    |x(t)| &\le \hat K |x(t_0)| e^{-\hat\lambda (t-t_0+n_{(t_0,t]}^{\sigma_\nu})}\quad \text{ for }0\le t_0\le t.
  \end{align}
  Since $\bar N<\lambda/(Ke^{\lambda})$ by assumption, then $\hat\lambda > 0$. Since~(\ref{eq:pertlinsolexp}) holds for every $x(t_0)$, then~(\ref{eq:tmpbound}) must be true.
\section{Conclusions}

Results for iISS, ISS and ISS under small inputs were given for families of impulsive systems, based on global state weak linearization and where {\color{blue}the convergence given by stability is ensured not only as time elapses but also as the number of occurring impulses increases, as is standard in the context of hybrid systems}. It was shown that the results do not remain valid if only the {\color{blue}more usual form of stability for impulsive systems is considered}, unless constraints are placed on the impulse-time sequences. Local exponential stability results were also provided, based on a first-order approximation. The given results generalize and extend existing ones in several directions. Some important system classes, such as bilinear systems, {\color{blue}and some switched system stablity problems} are covered. The given results do not require the zero-input system to be linear.

\bibliographystyle{IEEEtran}
\bibliography{/home/hhaimo/latex/strings.bib,/home/hhaimo/latex/complete_v2.bib,/home/hhaimo/latex/Publications/hernan_v2.bib}

%

\end{document}